\documentclass[aps,prl,superscriptaddres,floatfix,twocolumn,noshowpacs]{revtex4}
\usepackage{amssymb,amsfonts,amsmath}
\usepackage{graphicx,psfrag,xspace}
\usepackage{color}
\usepackage{natbib}
\usepackage{dcolumn}
\usepackage{bm}
\usepackage{subfigure}
\usepackage{setspace}
\usepackage{url}
\usepackage{xcolor}

\renewcommand{\d}{{\rm d}}

\begin{document}




\author{E. A. Jagla}
\affiliation{Centro At\'omico Bariloche and Instituto Balseiro, 
Comisi\'on Nacional de Energ\'{\i}a At\'omica, (8400) Bariloche, Argentina}  

\author{Fran\c cois P. Landes  and Alberto Rosso}
\affiliation{Laboratoire de Physique Th\' eorique et Mod\` eles Statistiques (UMR CNRS 8626), Universit\' e Paris-Sud, Orsay, France}


\title{Viscoelastic Effects in Avalanche Dynamics: A Key to Earthquake Statistics}

\begin{abstract} 
In many complex systems a continuous input of energy over time can be suddenly relaxed in the form of avalanches. Conventional avalanche models disregard the possibility of internal dynamical effects in the inter-avalanche periods, and thus miss basic features observed in some real systems. We address this issue by studying a model with viscoelastic relaxation, showing how coherent oscillations of the stress field can emerge spontaneously. Remarkably, these oscillations generate avalanche patterns that are similar to those observed in seismic phenomena.
\end{abstract}




\maketitle  


The driven dynamics of heterogeneous systems often proceeds by random jumps called avalanches, which display scale-free statistics. This critical out-of-equilibrium behaviour emerges from the competition between internal elastic interactions and interactions with heterogeneities, and is understood in the framework of the depinning transition \cite{Fisher1998, Kardar1998}. 
Remarkably, one can often disregard the precise details of the microscopic dynamics when considering the large scale behavior. 
As a result, various phenomena such as Barkhausen noise in ferromagnets \cite{alessandro1990domain, Zapperi1998, Durin2006}, crack propagation in brittle materials \cite{Alava2006, Bonamy2008, bonamy2011failure}  or wetting fronts moving on rough substrates  \cite{Rosso2002b, Moulinet2004, LeDoussal2009a} may display	similar avalanche statistics.

In this description of avalanches a trivial dynamics in the inter-avalanche periods is usually assumed \cite{Fisher1998, Sethna2001}.
However, the inclusion of viscoelastic effects with their own characteristic time scales brings about novel dynamical features. The existence of this kind of relaxation may have drastic consequences on the macroscopic behaviour of the system, as in the context of friction where it generates the time increase of static friction during the contact between two surfaces at rest \cite{Dieterich1972, Marone1998}. 	
Here we show how these relaxation processes generically induce a novel avalanche dynamics characterized by new critical exponents and bursts of aftershocks strongly correlated in time and space.
Due to its simplicity, the model allows for analytic treatment in mean field, and for extensive numerical simulations in finite dimensions.
Our main observations are twofold. 
First, in mean field the time scale of viscoelastic relaxation is associated with a dynamical instability, which we prove to be responsible for periodic oscillations of the stress in the entire system. 
This instability, named {\em avalanche oscillator},  was observed in numerical simulations and experiments of compression of Nickel micro crystals \cite{Papanikolaou2012}. 
Note that viscoelastic interactions are also at the root of the {\em hysteretic depinning} emerging in mean field periodic systems like vortex lattice or charge density waves \cite{Marchetti2000, Marchetti2005}.
Second, in two dimensions the global oscillations found in mean field remain coherent only on small regions. 
In each region the oscillations of the local stress have roughly the same amplitude and period but different phases, so that at a given time the stress map has a terraced structure.
 

We claim that the relaxation processes studied in our model are essential to capture the basic features of seismic dynamics. 
In particular, the viscoelastic time scale is the one involved in the aftershock phenomenon \cite{Dieterich1978, Scholz2002, Ben-Zion2008}. 
Moreover the oscillations of the stress field explain the quasi-periodic time recurrence of earthquakes that emerges from the data analysis of  the seismic activity in some geographical areas \cite{Barbot2012, Ben-Zion2003}.
Finally we show that in two dimensions, viscoelastic relaxation produces an increase in the exponent of the avalanche size distribution compatible with the Gutenberg-Richter law, and the aftershock spatial correlations obtained have strong similarities with the {\em migration effect} observed in real earthquakes \cite{Peng2009}.

\begin{figure}
\includegraphics{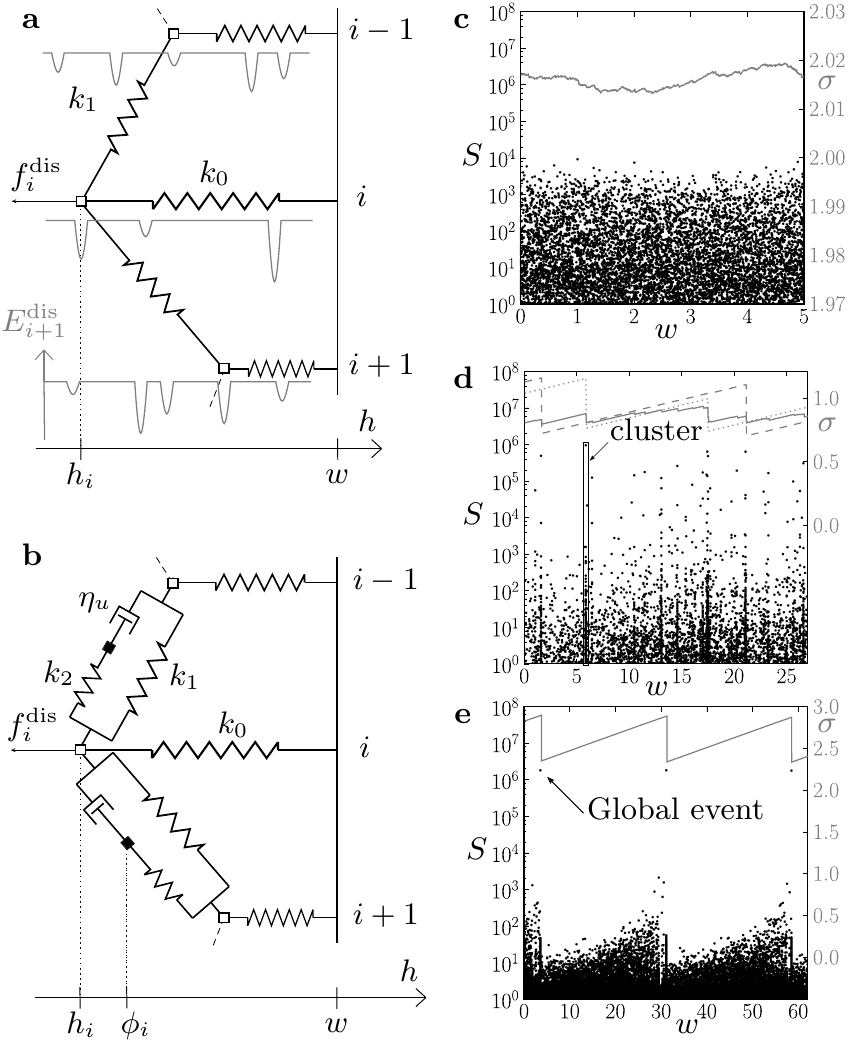}
\caption{\label{Fig1_V0}
Sketch of the models for $d=1$.  \textbf{(a)}
Conventional depinning model: disorder and elastic interactions acting on the blocks,
located in $h_i, h_{i+1}$, etc. The disordered force derives from the pinning potential (grey): $f_i^\text{dis}= -\partial E^\text{dis}_i(h_i)/\partial h_i$. 
\textbf{(b)} Depinning with relaxation: we introduce dashpots with relaxation
constant  $\eta_u$ and springs of stiffness $k_2$. 
Numerical results showing sequences of avalanches sizes $S$ and stress (grey) as a function of drive $w$ for: the elastic depinning model in $d=2$ \textbf{(c)}, the depinning with
relaxation in $d=2$ \textbf{(d)} (dashed and dotted grey lines correspond to local stress in two distant regions), and in mean field \textbf{(e)}.
}
\end{figure}

\textit{The models. --}
Our model with relaxation is constructed upon 
the paradigmatic model of avalanche dynamics, describing
the depinning of a $d$-dimensional elastic interface moving inside a $d+1$ dimensional space \cite{Fisher1998}. 
In this model, the interface consists of a collection of blocks (see Fig. \ref{Fig1_V0}a) obeying the equation of motion:
\begin{align}
\eta \partial_t h_i =  k_0 (w-h_i) + f_i^\text{dis}(h_i)  + k_1 \Delta h_i  \label{depinning}
\end{align}
where $(i, h_i)$ is the $d+1$-dimensional coordinate of the block and $\eta$ is the viscosity of the medium.
Each block feels elastic interactions via the (discrete) Laplacian term $k_1 (\Delta h)_i = k_1\sum_{<ij>} (h_j-h_i)$ (summation is restricted to nearest neighbors of $i$), disorder via $f_i^\text{dis}(h_i)$ and is driven towards the position $w = V_0 t$ via springs of elasticity $k_0$.
The force per unit area applied by the drive, namely the stress is given by $\sigma = k_0(w-\overline h)$, 
where $\overline h = \frac1N \sum_i h_i$ is the mean value of interface height ($N=L^d$).
The slow increase of $w$ over time induces an augmentation of the pulling force on each block. As a response, blocks typically adjust slightly their positions, but sometimes a block reaches a mechanically unstable state and moves far away from its position to a new local energy minimum. This can in turn destabilize neighbouring blocks, thus triggering an avalanche event that we characterize by its size $S=N (\overline{h}_\text{after}- \overline{h}_\text{before})$, which is simply the volume swept by the interface during the event.
In  Fig. \ref{Fig1_V0}c we show the sizes $S$ of a sequence of avalanches obtained by driving $w$ quasi-statically ($V_0=0^+$). 
The sequence displays an almost  Poissonian behavior, in
the sense that both the sizes and the 
time occurrences of the events
are almost uncorrelated variables.
Moreover the stress is constant in time, with 
fluctuations due to finite system size.

Our modified model consists in replacing springs $k_1$ by viscoelastic elements, built using springs and dashpots as  depicted in Fig. \ref{Fig1_V0}b.
Dynamical equations become:
\begin{align}
\eta \partial_t h_i
&=  k_0 (w-h_i) + f_i^\text{dis}(h_i)+k_1 \Delta h _i + k_2 (\Delta h_i - u_i) \notag \\
\eta_u \partial_t u_i
&= k_2 (\Delta h_i - u_i)  ,  \label{2}
\end{align}
where the auxiliary variables $u_i$ depend on the elongation of the neighbouring dashpots: in one dimension this variable reads $u_i = (\phi_i - h_i) + (h_{i-1} - \phi_{i-1})$  (see Supplemental Material).
The relaxation constant $\eta_u$ sets a new characteristic time $\tau_u = \eta_u/k_2$, to be compared with the two scales:
(i) $\tau_D = \overline{z}/V_0$ which accounts for the slow increase of the external drive $w$ (where $\overline z$ is the typical microscopic disorder length scale, defined later),
(ii) $\tau = \eta/\max [k_0,k_1,k_2]$, which is the response time of the $h$ variables.
Essentially, ``main'' avalanches are triggered by the drive through $k_0$, whereas relaxation (via $k_2, \eta_u$) triggers additional events on a time scale of order $\tau_u$: the aftershocks.

In our analysis, we assume that the three scales are well separated, namely $\tau \ll \tau_u \ll \tau_D$ (i.e. $\eta \ll \eta_u$). Hence, on the time scale $\tau$ 
the $u_i$'s are constant in time and the dynamics is exactly the same as for the depinning model with elastic constant $k_1+k_2$. 
However, after an avalanche, and in a time scale $\tau_u \gg \tau $, $h_i$'s are pinned (due to the narrow wells approximation, see below) and  $u_i$'s relax exponentially:
\begin{align}
u_i(t)=  \Delta h_i+ (u_i(t_0) - \Delta h_i ) e^{-(t-t_0)k_2/\eta_u} , \quad \forall i,
\label{fidet}
\end{align}
where $t_0$ is the time at which the last avalanche occurred. 
The effect of relaxation is to suppress the term $k_2 (\Delta h _i- u_i)$ in  Eq. (\ref{2}), so that some blocks may become unstable. This triggers secondary avalanches in the system, 
identified with aftershocks in the seismic context. 
Aftershocks occur without any additional driving: the ensemble of events that occur at a given  $w$ will be called a {\em cluster}  (Fig. \ref{Fig1_V0}d).
When $u_i= (\Delta h)_i$  $\forall i$, the system is {\em fully relaxed} and new instabilities can only be triggered by an increase of $w$.
Note that the fully relaxed configuration corresponds to a stable configuration of the depinning model with same disorder realization and elastic constants $k_0, k_1$.  

We mostly have in mind the $d=2$ case, which is the one realized in the seismic context. We consider local elastic interactions for implementation convenience. Realistic long range interactions within the faults (induced by the three dimensional nature of the plates) may have effects on the results we present, that are difficult to assess without a 
full numerical simulation. This remains as a prospect for future work.

\begin{figure}
\includegraphics[width=8.6cm]{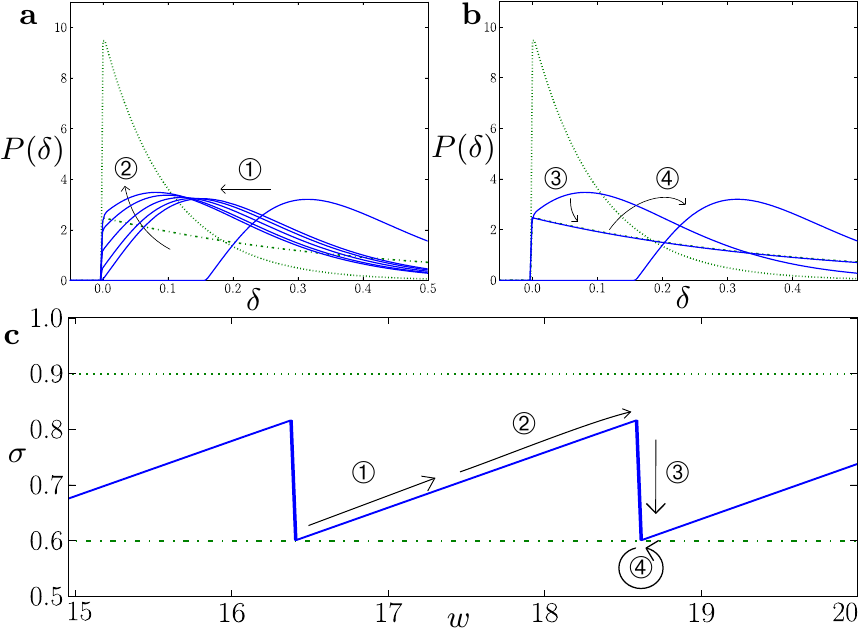}  
\caption{\label{FigPdelta}
Evolution of $P(\delta)$ (solid line) and the stress $\sigma= k_0(w - \overline{h})$ 
(lower panel) computed from direct integration of the evolution equations. 
($1$) driving without any avalanche, linearly increasing stress; 
($2$) driving with elastic-depinning avalanches, slower stress increase.
($3$) global event: $P(\delta)$ collapses to the depinning fixed point $Q(\delta, k_1+k_2)$ (lower dashed curve) and the stress drops to $\sigma(k_1+k_2)$ (lower dashed line). 
($4$) relaxation closes the cycle back to stage ($1$) without altering average stress. 
}	
\end{figure}

\textit{The narrow wells approximation. --}
To efficiently study Eqs. (\ref{depinning}, \ref{2}), we adopt the so-called ``narrow wells'' approximation. 
In this scheme, disorder is modeled as a collection of narrow pinning wells representing impurities (see Fig. \ref{Fig1_V0}a), with spacings $z$, distributed as $g(z)$ and with average $\overline{z}= \int_0^\infty z g(z) \d z$. The maximum value of the pinning force at each well is denoted $f^{\text{th}}$ 
(see the Supplemental Material).
Thanks to this choice, the system state can be reduced to a variable $\delta_i$. 
In the depinning model:
\begin{align}
\delta_i \equiv f_i^{\text{th}} - k_0 (w-h_i) - k_1 \Delta h _i.
\label{fth1}
\end{align}
As soon as $\delta_i \leq 0$, this site becomes unstable: an avalanche is triggered. The  avalanche evolution translates into simple rules for $\delta_i$'s. 
When $\delta_i>0, \forall i$, all blocks are stable and the avalanche is exhausted. Driving then follows until a new event is triggered.

In our model, $\delta_i$'s read:
 \begin{align}
\delta_i = f_i^{\text{th}} - k_0 (w-h_i) - k_1\Delta h _i - k_2(\Delta h _i- u_i).
\label{fth2}
\end{align}
The dynamics proceeds as before, with $u_i$'s kept constant during avalanches.
When the avalanche is exhausted, a slow relaxation of $u_i$ takes place (Eq. (\ref{2})).
This evolution can decrease $\delta_i$'s and thus trigger aftershocks.


\textit{Mean Field analysis. --}
We analyze the mean field, fully connected model, which corresponds to replacing $\Delta h_i$ with $\overline{h}-h_i$ in Eqs. (\ref{fth1}), (\ref{fth2}). 
In this case, all sites are equivalent and the $\delta_i$'s are independent and identically distributed variables,  characterized by the probability distribution $P_w(\delta)$ which in general depends on the initial condition $P_0(\delta)$ and on $w$.
In the Supplemental Material, we obtain the evolution of $P_w(\delta)$ under an infinitesimal increase in $w$ for both models, for $f_i^{\text{th}}=$ const. 

For the elastic depinning model, we show that this evolution has a fixed point reached within a finite increase in $w$
, at which $P_w(\delta)$ is given by the function:
\begin{align}
Q(\delta, k_1) = \frac{  1 - G(\frac{\delta}{k_0+k_1})}{\overline{z} (k_0+k_1)}   ,   \label{pdelta}
\end{align}
where $G(z)\equiv\int_0^z d z' g(z')$.
This indicates that the large time dynamics is stationary, and that 
the applied stress in the system is constant in time: $\sigma(k_1) \equiv f^ \text{th} - \overline{\delta}(k_1)$.
Further analysis shows that as long as $P_w(0) < (\overline{z} k_1)^{-1}$ the system displays avalanches bounded by a system-size independent cutoff: $S_\text{max} = ( 1- P_w(0) \overline{z} k_1 )^{-2}$. 
For example, at the fixed point (\ref{pdelta}) we have  $P_w(0)=Q(0,k_1)=1/\overline z (k_0+k_1)$, so that
$S_\text{max} = (\frac{k_0+k_1}{k_0})^2$.
However if $P_w(0) \geq (\overline{z} k_1)^{-1}$ the system becomes unstable, with a {\em global event}, that involves a finite fraction of the system.

For the model with relaxation, the evolution of $P_w(\delta)$ is non stationary and displays oscillations in time.
Under a small increase in $w$, two dynamical regimes are observed.
On  short times  $(t \simeq \tau )$, sites that become unstable move following the rules of a rigid elastic interface, with stiffness $k_1+k_2$.
On longer times  $(t\simeq \tau_u)$, during relaxation, the interface becomes more flexible (stiffness $k_1$), thus evolving towards the fixed point $Q(\delta,k_1)$ (\textit{stages} $1$ and $2$ in Fig. \ref{FigPdelta}).
However when $P_w(0)$ becomes larger than $1/\overline{z} (k_1+k_2)$, the rigid interface is unstable so that a single global avalanche drives $P_w(\delta)$ to the rigid fixed point $Q(\delta,k_1+k_2)$ (\textit{stage} $3$ in Fig. \ref{FigPdelta}). 
Finally, this state is altered by relaxation and a new cycle starts (\textit{stage}  $4$). 

Note that 
cyclic behaviour is independent of the 
details of the mean field model:
e.g.  Fig. \ref{Fig1_V0}e corresponds to the case of randomly distributed thresholds $f^\text{th}_i$.
The avalanche dynamics is different and displays aftershocks, but global events and stress oscillations are also present.

A similar visco-elastic model (with $k_1=0$, periodic disorder and under constant force $F$) was originally introduced \cite{Marchetti2000, Marchetti2005, LeDoussal2008} to model the hysteretic depinning observed in vortex lattices and charge density waves. 
In the fully connected approximation a self-consistent calculation pointed out that the average velocity is multi-valued, yielding  hysteretic behavior in a wide range of external stress.
This hysteresis echoes with the 
oscillations in  
Fig. 2. Yet, in two dimensions, we 
see that constant velocity driving produces stress distributions that are not uniform, nor constant, and we 
get qualitative new results that were not obtained 
in the constant applied force case studied in \cite{Marchetti2000, Marchetti2005, LeDoussal2008}).



\begin{figure}
\includegraphics{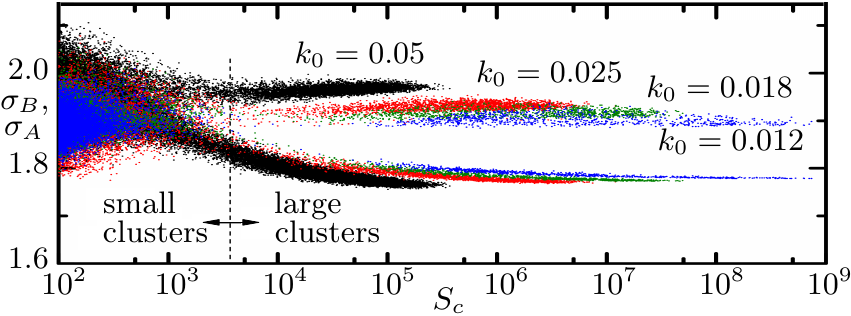}
\caption{
The local stress restricted to the cluster area, just before (up, $\sigma_B$) and just after (bottom, $\sigma_A$) it takes place, as a function of the cluster size $S_C$. 
Thus, in e.g. a compression experiment, one expects the average local variation of stress to vanish (with undefined values of $\sigma_{B,A}$) for small avalanches, and to saturate to a constant (nonzero) value (with well defined  values for $\sigma_{B,A}$) for large avalanches.
}
\label{sasb}
\end{figure}

\begin{figure}
\includegraphics{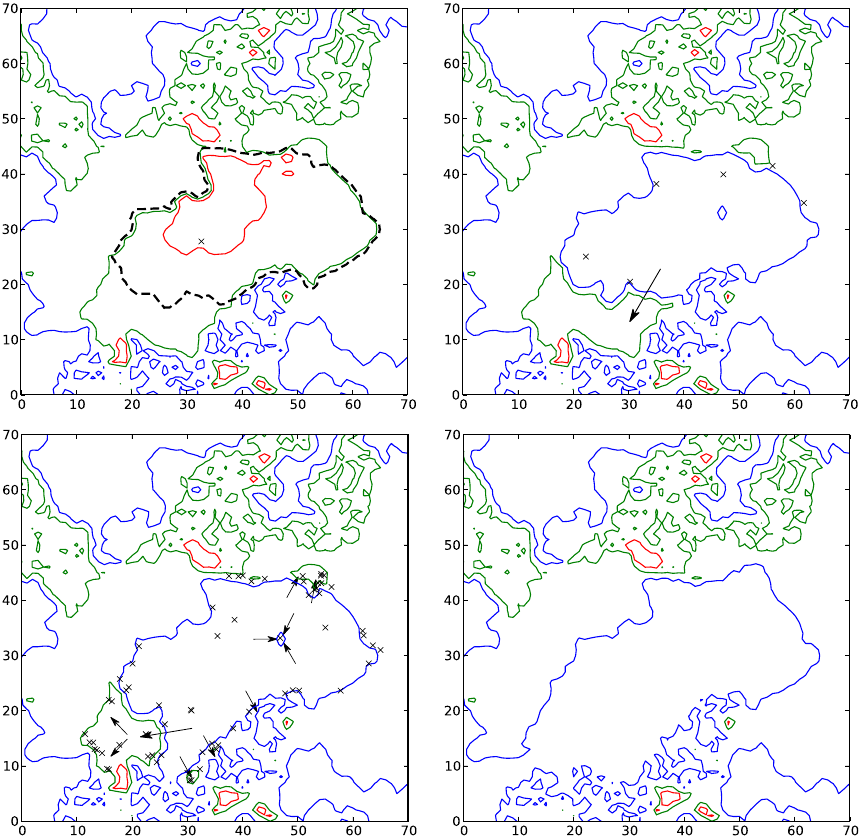}
\caption{\label{figBrokenZone}
Stress map in $d=2$. Colors indicate stress levels, from high (red) to low (blue).
Upper left: stress map just before a large event, with the unstable region highlighted by a dashed line.
From left to right and top to bottom: expansion of the affected area is seen to mainly spread (black arrows) around the initial main shock and the subsequent aftershocks (small crosses). Affected regions have low chance to witness new large events, due to the low value of the local stress. 
}
\end{figure}	


\textit{Two dimensional results. --}
For $d=2$ we must rely on the numerical implementation of Eqs. (\ref{fth2}, \ref{fidet}) via an efficient method originally developed in \cite{Grassberger1994a} (see Supplemental Material).
In Fig. \ref{Fig1_V0}d, we see a clear distribution of events in clusters of main shocks and aftershocks, as in actual seismicity (where, indeed, any single cluster spans a finite $w$ interval, due to the incomplete separation of time scales).
The periodicity of the mean field (Fig. \ref{Fig1_V0}e) has now disappeared.

Nevertheless, a careful analysis of the $d=2$ model shows an interesting reminiscence of the mean field behaviour.
In Fig. \ref{sasb} we compute for each cluster the local stress restricted to the cluster area, just before ($\sigma_B$) and just after ($\sigma_A$) it takes place 
(the same analysis for events instead of clusters yields the same results).
Small clusters show broad distributions of $\sigma_B$ and $\sigma_A$, similar to what would be observed for the depinning case.
However for large clusters both distributions become very narrow: $\sigma_B$ sets to a value that we denote $\sigma_{\max}$, and $\sigma_A$ sets to $\sigma_{\min}$.
This is the fingerprint of the mean field behaviour, suggesting a large scale description of the $d=2$ interface as a terraced structure.
Indeed, we observe (Fig. \ref{figBrokenZone}) that different parts of the system have different values of the stress, which range from  $\sigma_{\min}$ to $\sigma_{\max}$.
In analogy with mean field, when the stress of a region reaches a value $\sim \sigma_{\max}$, it gets destabilized and the whole region collapses to $\sigma_{\min}$. 
In fact the evolution of the local stress associated with a small patch of the interface is non stationary, and shows an almost periodic oscillation between $\sigma_{\min}$ and $\sigma_{\max}$ (Fig. \ref{Fig1_V0}d, dashed and dotted lines). However this oscillation is not synchronized among different patches, so the system does not display a global oscillation.
It is remarkable that 
the width of the distribution of the local stress ($\sim \sigma_{\max}- \sigma_{\min}$) remains finite  when $k_0\to 0 $,
while in the depinning model \cite{Rosso2012}, it vanishes as $k_0^{1-\zeta/2}$ for very small $k_0$ ($\zeta$ is the roughness exponent of the interface which is found to be smaller than $2$).	
Moreover our model supports the idea that seismic activity in some geographical regions displays quasi-periodicity (the so-called seismic cycle \cite{Scholz2002}).
This periodicity was recently studied in the context of  micro-crystals deformation  \cite{Papanikolaou2012}, where 
it was 
named ``avalanche oscillator". 
Similar kind of oscillations were also observed in models with relaxation \cite{Rosso2012}, granular materials \cite{Dahmen2011} and molecular dynamics of visco-elastic disordered systems \cite{Salerno2013}.

A second important feature 
is the spatial distribution of aftershocks in a given cluster (see  Fig. \ref{figBrokenZone}).
After a main shock, many aftershocks follow, extending the slip area.
The small ones (not indicated) are rather uniformly distributed inside the slip region;
while the epicentres of the large ones typically occur at the border, extending the slip region.  
Field observations report this effect as  ``aftershock migration" \cite{Peng2009}.

As a third point, we discuss the size distribution of the avalanches in 2D, presented in Fig. 2 of the Supplemental Material. 
We find a consistent power law decay in all the range that we have been able to explore (at least in a size range of $10^7$) with an {\em anomalous} exponent $\kappa\simeq 1.7-1.8$.
This is quite remarkable, given that in all conventional avalanche models like depinning or directed percolation, this exponent is always smaller than $1.5$, which corresponds to the mean field limit \cite{LeDoussal2009, Dobrinevski2012}. 
In particular in the 2D depinning case one measures $\kappa \simeq 1.27$ \cite{Rosso2009}.  
Our result can be compared with the value for actual earthquakes where where $\kappa$ values in the range $\simeq 1.7\pm 0.2$ are reported \footnote{Note that historically the magnitude of an earthquake is defined as $M=(2/3) \log_{10} S$. The Gutenberg-Richter law (GR) states that $N(M) \sim 10^-{bM}$, with reported values of $b$ in the range $1\pm 0.25$, so that from the definition $N(S) \sim S{-\kappa}$, we obtain approximately $\kappa= 1+2b/3\simeq 1.7 \pm 0.2$.}.
However, since we do not consider realistic long range elastic interactions, this coincidence has to be taken with caution.
We also note that a justification for the value of the GR exponent has been given recently using a forest fire model analogy \cite{Jagla2013}. 
It is worth mentioning that arguments in \cite{Jagla2013}
build upon a model that has a terraced structure of the interface compatible with the
one we find here.


\textit{Conclusions. --}
Internal relaxation plays crucial roles in the dynamics of sliding objects, and becomes particularly important at large scales, relevant to seismic phenomena. 
The dynamics of our model
shows a strong tendency to become non-stationary. 
This tendency is manifest in mean field, where sliding proceeds as a sequence of
global and periodic stick-slips. 
In $d=2$, we provide
numerical evidence that periodic stick-slips occur locally,
without global synchronization 
across 
the system.
There, our predictions mainly deal with spatial properties of events, thus demanding high spatial resolution in experiments. 

\begin{acknowledgments}

We thank Shamik Gupta and Mikhail B. Zvonarev for useful discussions.
We acknowledge support from the France-Argentina MINCYT-ECOS A12E05. E.A.J. is financially 
supported by CONICET (Argentina). Partial support from grant PICT-2012-3032 
(ANPCyT, Argentina) is also acknowledged.

\end{acknowledgments}


\newpage


\begin{center}
{\LARGE \bf Supplemental Material}\\
\end{center}

\section{1. Depinning with relaxation: derivation of the equations and dynamical properties}

\subsection{1.1 Derivation of the equations of motion }


The depinning model with relaxation corresponds to the mechanical circuit sketched in Fig. 1b of the main text. We first describe the one-dimensional case.
The sample is decomposed in blocks of mass $m$, labeled $i$ and moving along horizontal rails $h_i$. 
The action of the dashpot is to resist the change in $\phi_i - h_i$ via viscous friction, with a resulting force on $h_i$ given by $\eta_u  \partial_t (\phi_i-h_i)$.
The blocks move in a medium with an effective viscosity $\eta$, we will study the overdamped regime $m \partial_t^2 h_i \ll \eta\partial_t h_i$.
As each block is described by two degrees of freedom $h_i$ and $\phi_i$, its time evolution is governed by two equations.
The first equation comes from the force balance on $h_i$:
\begin{align}
\eta \partial_t h_i 
=&  f^\text{dis}_i(h_i) + k_0 (w-h_i) + k_1(h_{i+1} - h_i)   \\
&+ k_1 (h_{i-1} -h_i) + \eta_u \partial_t(\phi_i -h_i) + k_2 (\phi_{i-1}-h_i) \notag
\label{viscoqEW1}
\end{align}
The second equation is derived from the force balance on $\phi_i$: 
\begin{align}
0= k_2(h_{i+1} - \phi_i) + \eta_u \partial_t(h_i - \phi_i)
\end{align}
where we assumed that the internal degree of freedom $\phi_i$ has no mass. 
We inject this second equation into the first, and subtract the force balance on $\phi_{i-1}$ to the second equation.
It is convenient to let the Laplacian term $k_2 (h_{i+1}- 2h_i+h_{i-1})$ appear by defining the variable $u_i =\phi_i - h_i + h_{i-1} - \phi_{i-1}$:
\begin{align}
\eta \partial_t h_i =& f^\text{dis}_i(h_i) + k_0 (w-h_i) + k_1(h_{i+1} -2 h_i + h_{i-1} )  \notag \\
&+k_2 (h_{i+1}- 2h_i+h_{i-1}) - k_2 u_i  \notag \\
\eta_u \partial_t u_i &= k_2 (h_{i+1}- 2h_i+h_{i-1}) - k_2 u_i
\end{align}


To generalize this to higher dimensions (on a square lattice), one simply has to connect each block $h_i$ to its neighbours via symmetrically arranged viscoelastic elements.
The equations obtained are exactly the same, with the label $i$ now referring to $d$-dimensional space, the $d=1$ discrete Laplacian replaced with the $d$-dimensional Laplacian denoted $\Delta$, and the $u_i$ variable redefined as:
\begin{align}
u_i = \sum_{j=1}^ {d}( \phi_j - h_j) + \sum_{j'=d+1 } ^ {2d} (h_{j'} - \phi_{j'}) \label{dDgeneral}
\end{align}
In this way, one obtains Eq. (2) of the main text.

It is worth to notice that the dynamics of the model without relaxation satisfies the Middleton theorem  \cite{middleton1992asymptotic} that guarantees that the interface moves only forward.
However, in presence of viscoelastic elements the Middleton theorem does not apply, and  backward movements of the interface are possible. 
Fortunately, these movements are not frequent, due to the biased driving term $k_0(w-h)$ and we observed numerically that the real dynamics yields the same statistical results as the dynamics that allows only forward movements. 
Thus, we restrain the dynamics to forward movements.

\subsection{1.2 The narrow wells approximation }

To efficiently study Eqs. (1) and (2) of the main text, we adopted the so-called ``narrow wells'' approximation, which does not affect the universal properties of the dynamics, and can be extended to the model with relaxation. 
This approach greatly simplifies the numerical simulations and allows an analytical analysis in the mean field case.
In this scheme, the disorder is modeled as a collection of narrow pinning wells representing impurities (see Fig. 1a). 
Along the $h$ direction, the pinning wells are separated by random intervals (spacings) $z$ with distribution $g(z)$ and mean length $\overline{z}= \int_0^\infty z g(z) \d z$. 
A natural choice for $g(z)$ is the exponential law, which corresponds to the case where impurities are uncorrelated in space. 
If the spatial extension of the wells is negligible compared to $\overline z$, we can safely assume that each block is 
 located in one of those wells, so that its coordinate $h_i$ evolves only by discrete jumps $z$.
To exit from a well, a block needs to be pulled by a force larger than a threshold $f^{\text{th}}_i$ related to the well's depth.
Within this approximation, the continuous dynamics of the blocks can be re-written in terms of the variable $\delta_i$, which  measures the remaining stability range of the block $i$. In the depinning model it reads:
\begin{align}
\delta_i \equiv f_i^{\text{th}} - k_0 (w-h_i) - k_1 \Delta h _i.
\label{fth1}
\end{align}
Increasing the load $w$, all $\delta_i$'s decrease, until a block becomes unstable ($\delta_i=0$) and  moves to the next pinning well ($h_i \mapsto h_i + z$), characterized by a new random threshold $f_i^\text{th}$.
The unstable block can be the seed of an avalanche because its motion produces a drop $k_1 z$ in the variables $\delta$ of the neighbouring blocks.
The avalanche event is exhausted when all blocks are stable.

\section{2. Mean field analysis}

In general, the mean field limit can be studied using models which are much simpler than their finite dimension counterpart.
For example, the mean field depinning can be mapped \cite{Zapperi1998} onto the  problem of a single particle driven in a Brownian force landscape, the so-called ABBM model \cite{alessandro1990domain}.
Many results on the avalanche statistics of the mean field interface can be obtained from this latter model \cite{LeDoussal2009, Dobrinevski2012}.
Unfortunately, such a mapping does not hold in presence of relaxation (recently, a version of the ABBM model ``with retardation'' was studied \cite{Dobrinevski2013}, displaying aftershocks but no oscillatory behavior).
A different strategy, which can be generalized to that case, is to consider the fully connected model, where each site interacts with all the others.

In the fully connected model, all sites are equivalent and the $\delta_i$'s are independent and identically distributed variables,  characterized by their probability distribution $P_w(\delta)$ which in general depends on the initial condition $P_0(\delta)$ and on the value of $w$.
The aim of this section is to write down the evolution equation for $P_w(\delta)$ when $w$ increases.
We will show that in the depinning case the distribution reaches a stationary form, while the viscoelastic depinning displays a periodic solution.

%

\subsection{2.1 Reference material for the conventional depinning model}

By replacing the local term $(\Delta h)_i$ in Eq. (4) of main text with its fully connected version $\overline h-h_i$, we obtain:
\begin{align}
\delta_i= f^\text{th} - k_0(w-h_i) - k_1(\overline{h} -h_i) 
\end{align}
Let us set the threshold force $f^{\text{th}}$  to be constant, 
a choice that, for the mean field analysis, does not alter the results.

When the external driving is increased by a small positive quantity $\d w$, the distribution evolves from its initial shape $P_w(\delta)$, to a new shape $P_{w+\d w}(\delta)$. 
In order to compute the latter, it is useful to decompose the dynamical evolution in different steps. 
In a first step, the center of the parabolic potential moves from $w$ to $w + \d w$ and all $\delta_i$'s decrease by $d \delta = k_0 \d w$, moreover a fraction  $P_w(0) k_0  \d w $  of sites becomes unstable and moves to the next wells. 
The new $\delta_i$ are given by $z (k_1+k_0)$,  with $z$'s drawn from the distribution $g(z)$. 
This  writes:
\begin{align}
\frac{P_{{\rm step} 1} (\delta) - P_w(\delta)}{ k_0  \d w }
&=  \frac{\partial P_w}{\partial \delta} (\delta)  +  P_w(0) \frac{ g\left( \frac{\delta}{k_0+k_1}\right) }{ k_0+k_1}  \label{EqPdelta}
\end{align}
The redistribution of $\delta$'s changes $\overline{h}$ of a quantity $ P_w(0) \overline{z} k_1$, so that all blocks are subject to a shift in their $\delta_i$.
This can induce a second step which acts on $P_{{\rm step} 1}(\delta)$ exactly as ${\rm step} 1$ did on $P_w(\delta)$, but with a shift $d \delta = P_w(0) \overline{z} k_1$.
These steps go on until there are no more unstable sites, so that the distribution reaches the stable form $P_{w+ \d w }(\delta)$.
Let us remark that Eq.[\ref{EqPdelta}] has a fixed point $P_*(\delta)$ found when 
\begin{align}
 \frac{\partial P_*}{\partial \delta} (\delta)  +  P_*(0) \frac{ g\left( \frac{\delta}{k_0+k_1}\right) }{ k_0+k_1} = 0 .
 \end{align} 
 This equation can be easily integrated and $P_*(0) $ determined by the normalization condition.
This gives:
\begin{align}
P_* (\delta)  &= Q(\delta, k_1) = \frac{  1 - G(\frac{\delta}{k_0+k_1})}{\overline{z} (k_0+k_1)}   ,   \label{pdelta}
\end{align}
where $G(z)\equiv\int_0^z d z' g(z')$.
A stability analysis shows that the fixed point is attractive, so that any initial condition converges to it.
Moreover, it is possible to prove that for a given initial condition, there exists a finite $w_*$ at which the distribution reaches the fixed point and remains there for $w>w_*$.
This indicates that the large time dynamics is stationary, and that the applied stress in the system is constant in time:
\begin{align}
\overline{\sigma}(k_1) \equiv f^ \text{th} - \overline{\delta}(k_1).
\end{align}
This result becomes $\overline{\sigma}(k_1)=  f^ \text{th} - (k_0+k_1)\overline{z}$ for an exponentially distributed $z$.


For the depinning case, we can also compute the probability distribution of the avalanche sizes $N(S)$ in the fully connected approximation, for finite values of the parameters $k_0, k_1, \bar{z}$. Let us first consider the case where $g(z)=\delta(z-\overline{z})$.
For a finite system with $N$ sites, the typical configuration $\{\delta_i\}$ corresponds to a set of $N$ independent and identically distributed random variables drawn  from $P(\delta)$.
Let us sort the set: $ \delta_0 < \delta_1 < \dots <\delta_{N-1}$. 
When the system becomes unstable we have by definition $\delta_0=0$.
This site jumps to the next well at distance $\overline{z}$, so that all $\delta_i$'s are decreased by $\overline{z} k_1/N$. This will produce at least another jump if $\delta_1<  \overline{z} k_1/N$. 
More generally, the avalanche size $S$ corresponds to the first time  that the relation:
\begin{align}
\delta_{S-1} \leq   \frac{\overline{z} k_1}{N} S < \delta_S 
\end{align}
is fulfilled.

It is thus important to study the statistics of the  $\delta_i$ with $i \ll N$. 
Let us observe that when  $N$ is very large the distribution of these $\delta_i$'s can be approximated with a uniform distribution: $ P(\delta) \sim P(0)$.
Within this approximation, the spacings $X_i=\delta_{i+1}-\delta_i$ are independent exponential variables of mean $1/P(0)N$ and variance $1/	(P(0)N)^ 2$.
We conclude that the sequence $\delta_0, \dots, \delta_i$ is a random walk with diffusion constant $1/(P(0)N)^ 2$ and drift $1/(P(0)N)$.

The statistics of $S$ thus corresponds to the problem of first crossing with $0$ of  a random walk with diffusion constant $D=1/(P(0)N)^ 2$ and drift $d =  \frac{\bar{z}k_1}{N}- \frac{1}{P(0)N}$. For a positive drift, there is a finite probability that this random walk never crosses $0$, which corresponds to a {\em global} event.
For a negative drift, the time of zero crossing  is always finite, and has been computed for the Brownian motion in  \cite{Majumdar2002}.
The distribution of the avalanche sizes thus reads:
\begin{align}
N(S) &\sim  S^ {-3/2} e^ {-S/2S_\text{max}}  \notag\\
\text{with } &S_\text{max} = \frac{D}{d^ 2}  
= (1 - P(0)\overline{z}k_1 )^ {-2} \label{Smax}
\end{align}
where for simplicity we have neglected the short-scale regularization in the expression of $N(S)$. 
If now we replace the choice $g(z)=\delta(z-\overline{z})$ with a broader function $g(z)$, the results of Eq. [\ref{Smax}] still hold but  with a different diffusion constant.
Finally let us remark that the results we obtain here by focusing on $\delta_i$ coincide with the results obtained using the mapping to the ABBM model.


\subsection{2.2 Mean field analysis of the model with relaxation}

The fully connected approximation for the model with viscoelastic elements is obtained by replacing $(\Delta h)_i$ in Eq. (5) of main text  with $\overline h-h_i$, as for  the depinning model:
\begin{align}
\delta_i&= f^\text{th} - k_0(w-h_i) - (k_1+k_2)(\overline{h} -h_i) + k_2 u_i. \label{a}
\end{align}
It is useful to split $\delta$ in a {\em fast} part, $\delta^F$, and a {\em relaxed} one, $\delta^R$: 
\begin{align}
 \delta^F _i &= f^{\text{th}} - k_0 (w-h_i) - (k_1 + k_2) (\overline{h}-h_i)\notag  \\
  \delta^R _i &=   k_2 u_i . \label{defdeltaF} 
\end{align}
Under a small increase of $ \d w $, two dynamical regimes are observed.
On the short time scales $(t \simeq \tau )$ the dashpots are blocked, so that all $\delta^ R_i$'s remain constant while all $\delta^F_i$ are shifted by $k_0  \d w $. 
Sites that become unstable (i.e. with $\delta_i=\delta^F_i+ \delta^ R_i \leq 0$) move to their next pinning well, following the rules of the rigid depinning interface, with stiffness $k_1+k_2$.
On longer time scales $(t\simeq \tau_u)$ the dashpots relax, so that $\delta^R_i$'s slowly evolve and can possibly trigger new fast events called aftershocks. 
Within the approximation $f^\text{th} = \text{const}$, no aftershocks are observed.
This allows the blocks to fully relax after each event, so that the system's state just before any event is always fully relaxed ($u_i = \overline h - h_i$).
This corresponds to a relaxed $\delta^ R_i$:
\begin{align}
\delta^R_{i,\infty} = k_2 (\overline{h}-h_i)  = k_2   \frac{\overline{\delta^F}-\delta_i^F }{k_0+k_1+k_2} \label{fullyRelaxed}
\end{align} 
where the last equality is obtained by inverting Eqs. [\ref{defdeltaF}].
The effect of an event is to modify the distribution of $\delta^ F$ which, just before an event, is related to $P_w(\delta)$ (using Eq. [\ref{fullyRelaxed}]): 
\begin{equation}
\widetilde{P}_w(\delta^ F) = \kappa P_w ( \kappa (\delta^ F + \delta^{*} )  ),
\label{Ptilde}
\end{equation}
where $\delta^ *=k_2 \overline{\delta} / (k_0+k_1)$ and $\kappa = (k_0+k_1) / (k_0+k_1+k_2)$.
At the first step of the event, the unstable sites are those with $\delta^ F=-\delta^ *$, so that  $\widetilde{P}$ evolves from $\widetilde{P}_w$ to $\widetilde{P}_{\text{step}1}$ via:
\begin{align}
\frac{\widetilde{P}_{\text{step}1} - \widetilde{P}_w }{ k_0 \d w } 
= \frac{\partial \widetilde{P}_w}{\partial \delta^F} 
+   \frac{\widetilde{P}_w(-\delta^{*}) }{k_0+k_1+k_2} 
 g\left( \frac{\delta^F + \delta^{*} }{k_0 +k_1+k_2}\right)  \label{PdeltaEvolution}
\end{align}
this equation has a depinning-type fixed point $\widetilde{P}_*(\delta^F)= Q(\delta^ F+\delta^*, k_1+k_2)$, which translates for the variable $\delta$ in the fixed point $P_*(\delta)=Q(\delta,k_1)$ of the depinning model for the more flexible interface (with stiffness $k_1$).
We conclude that when the dynamics consists in avalanches of very few {\em steps} (here we assume just one), the distribution $P_w(\delta)$ converges to this attractive fixed point.
The dynamics actually consists in avalanches of very few steps (which are furthermore of decreasing amplitude), so that the distribution $P_w(\delta)$ indeed goes ``up'' towards this attractive fixed point.

This convergence, observed in Fig. 2a of the main text, stops when $P_w(\delta)$ overcomes the fixed point of the rigid interface at $\delta=0$, namely  $P_w(0) \geq (\overline{z}(k_1+k_2))^ {-1}$.
At this stage, a global avalanche is triggered in the system
\footnote{For avalanches that last for more than one step, the evolution of $P_w(\delta)$ can not be computed from Eq. [\ref{PdeltaEvolution}] because the system is not fully relaxed during the avalanche. 
Instead it is necessary to follow the evolution of the joint probability distribution $P(\delta^F, \delta^ R)$. 
The details of this numerical integration are given in the Supplemental Material Sec. 3.1}, and $P_w(\delta)$ jumps to the fixed point $Q(\delta, k_1+k_2)$ of the rigid interface (stage $3$).
Finally, at the end of this global event, relaxation takes the system far from this fixed point, and a new cycle starts (stage $4$).



\begin{figure}
\begin{center}
 \includegraphics{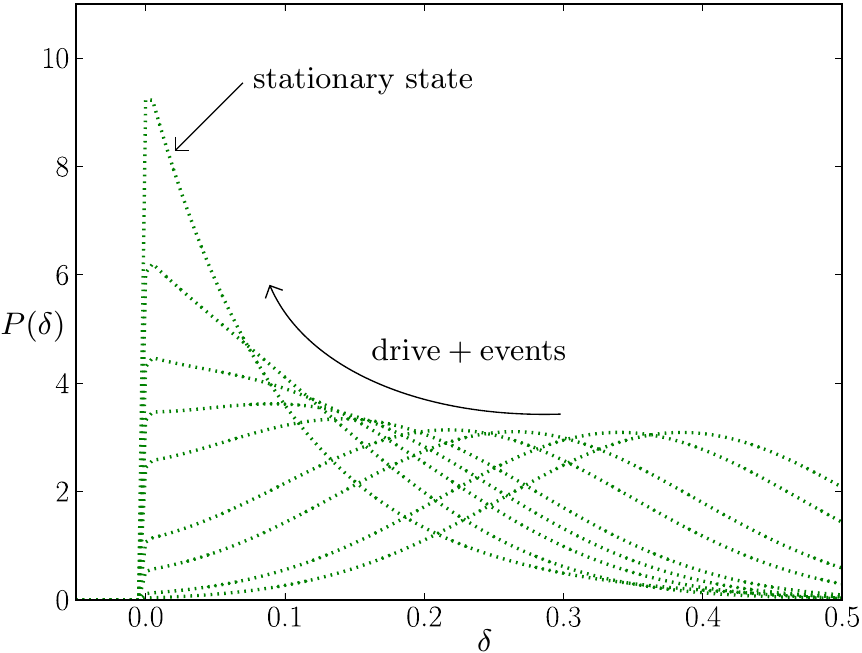}
  \caption{\label{convergence} The evolution of $P_w(\delta)$ for the quenched Edwards-Wilkinson model when $w$ is increased. The initial distribution is a Gaussian centred in $\delta=0.4$, with standard deviation $0.15$, and the weight at the left of $\delta=0$ cut. $P(\delta)$ quickly reaches its stationary form.}
\end{center}
\end{figure}

\section{3. Fully connected model: numerical integration of the evolution equations of $P(\delta)$}

\subsection{3.1 Numerical integration for the depinning model}

Let us discretize $P(\delta)$ with a bin of size $\varepsilon$. 
The distribution probability is then a vector $P_i$ (related to $P(\delta)$ by $P_i = P(\delta = \varepsilon i)$) which evolves with the following rules:
\begin{itemize}
\item {\em Driving process}:
We shift  $P_i$ of one bin: $P_i \leftarrow P_{i+1} $.
\item {\em Instability check}:
We compute the weight of unstable sites: $$P_{\text{inst}} = \varepsilon \sum_{i<0} P_i $$ 
If $P_{\text{inst}} > 0 $, we perform the {\em Avalanche process}. \\
Else we go back to the {\em Driving process}.
\item {\em Avalanche process}: it is composed by a stress drop and a stress shift.
\begin{itemize}
\item Stress drop: 
\begin{align}
P_{i\geq0}&\leftarrow P_{i}+  P_{\text{inst}} \frac{g\left(\varepsilon i/(k_0+k_1)\right)}{k_0+k_1}  \nonumber\\
P_{i<0}&\leftarrow 0  \nonumber
\end{align}
\item Stress shift: we shift $P_i$ of $n_{\text{shift}} = \text{Int} [\frac{\overline{z} k_1 P_{\text{inst}}}{\varepsilon}]$ bins.
\begin{align}
P_i \leftarrow P_{i+n_{\text{shift}}}  \nonumber
\end{align}
\end{itemize}
\noindent Then we perform the {\em Instability check}.
\end{itemize}
This algorithm converges very quickly from any initial configuration to $P_*(\delta)$ for any choice of $g(z)$, see Fig. \ref{convergence} of the Supplemental Material.

\subsection{3.2 Numerical integration for the model with relaxation}\label{sectionAlgov3}

Analogously to the previous case, we discretize $P(\delta^F, \delta^R)$ with a bin $\varepsilon$.
The distribution probability is then a matrix $P_{i,j}$ where $P_{i,j} = P(\delta^F = \varepsilon i, \delta^R = \varepsilon j)$. 
The matrix evolves with the following rules:
\begin{itemize}
\item {\em Driving process}:
We shift $P_{i,j}$ of one bin: \\
 $P_{i,j} \leftarrow P_{i+1,j} $.
\item {\em Instability check}:
We compute $P_0=\sum_{i=-j} P_{i,j}$. If $P_0 \overline{z} (k_1+k_2) \geq 1$, we perform the  {\em Driving process}.

We compute the weight of unstable sites: $$P_{\text{inst}} =  \varepsilon  \sum_{(i+j) < 0} P_{i,j}$$ \\
If $P_{\text{inst}} > \frac{\varepsilon}{\overline{z} (k_1+k_2)} \frac{1}{100} $, we perform the {\em Avalanche process}. \\
Else we perform the {\em Relaxation process}.
\item {\em Avalanche process}: it is composed by a stress drop and a stress shift.
\begin{itemize}
\item Stress drop: $\forall (i,j)$, \\
if $i+j\geq0$ :
\begin{align*}
P_{i,j} \leftarrow  P_{i,j} + \frac{\varepsilon}{\kappa} \left( \sum_{i'|(i'+j<0)} P_{i',j}\right)     g\left(\frac{\varepsilon (i + j)}{\kappa}\right) 
\end{align*}
if $i+j<0$ : 
\begin{align*}
P_{i,j}& \leftarrow 0,  
\end{align*}
where $\kappa=k_0+k_1+k_2$.
\item Stress shift: we shift $P_{i,j}$ of a fraction of bin: \\
$r = \min (1, \frac{\overline{z} (k_1+k_2) P_{\text{inst}}}{\varepsilon})$,
\begin{align}
P_{i,j} \leftarrow  P_{i,j} +  \left( P_{i+1,j}-P_{i,j} \right) r  \nonumber
\end{align}
\end{itemize}
\noindent Then we perform the {\em Instability check}.
\item {\em Relaxation process}:
We compute $j_{\infty}(i) $, the  single bin associated to $\delta^R_{i,\infty}= j_{\infty}(i) \varepsilon$ as\footnote{It is numerically more stable to associate $\delta^R(i,\infty)$ with two bins, $j_{\infty}(i)$ and $j_{\infty}(i)+1$. The contribution  $\sum _j P_{i,j}$ is split in the two bins using a linear interpolation.}
$$j_{\infty}(i)=  \text{Int} \left(  k_2  \frac{- i +  \sum_{i',j} i' P(i',j)  }{\kappa} \right) $$ 
so that the relaxation corresponds to:
\begin{align}
&P_{i,j_\infty(i)} \leftarrow  \sum _j P_{i,j}  \nonumber\\
&P_{i,j \ne j_\infty(i)} \leftarrow 0 \nonumber
\end{align}
Then we perform the {\em driving process}.
\end{itemize}

This algorithm integrates the fully connected version of the viscoelastic model, and produces the results shown in Fig. 2 of the main text.

\section{4. Two dimensional case: details on the numerical integration procedure}

We provide here details on the integration of the dynamic equations of the viscoelastic model. Our starting point is the set of equations (2) of the main text:
\begin{align}
\eta \partial_t h_i
&=  k_0 (w-h_i) + f_i^\text{dis}(h_i)+k_1 \Delta h _i + k_2 (\Delta h_i - u_i) \label{1}\\
\eta_u \partial_t u_i
&= k_2 (\Delta h_i - u_i)\label{2}
\end{align}
with $w=V_0t$.
For the numerical work, it is convenient to introduce variables $F_i$ and $G_i$, defined as:
\begin{align}
F_i&\equiv k_2 (\Delta h_i - u_i), \\
G_i&\equiv k_1 (\Delta h_i)+k_0(w-h_i).
\end{align}
Using $F_i$ and $G_i$, the model equations can be written as
\begin{align}
\eta \partial_t h_i
&=  f_i^\text{dis}(h_i)+G_i + F_i\label{hdot}\\
\eta_u \partial_t F_i +k_2 F_i&= \eta_u k_2 (\Delta \partial_t h)_i.
\label{fdot}  
\end{align}
It is thus clear that $G_i$ represents the force onto $h_i$ exerted through $k_1$ and $k_0$ springs, whereas $F_i$ is the force coming from the branches that contain the dashpots and $k_2$ springs.

We work in the case in which temporal scales are well separated: $\tau \ll \tau_u \ll \tau_D$. 
This corresponds to $\eta \ll \eta_u \ll \overline z k_0 /V_0$. 
As discussed in the main text, within the narrow well approximation the actual integration of Eqs. [\ref{hdot}] and [\ref{fdot}] does not need a continuous time algorithm, but can be presented in the form of a discrete set of rules.
From a relaxed configuration with $F_i=0$ at time $t$, the load increase triggers a new instability of Eq. [\ref{hdot}] when the total force from the springs, here $G_i$, reaches $f_i^{th}$, and this occurs after a time interval:
\begin{equation}
\delta t=\min_i \left (\frac{f_i^{th}-G_i}{k_0 V_0}\right )
\label{dtext}
\end{equation}
Thus at time $t+\delta t$ an avalanche starts at position $i$, producing the advance of $h_i$ to the next potential well $h_i \leftarrow h_i +z$, and a corresponding rearrangement of the forces according to (in two dimensions):
\begin{align}
F_i&\leftarrow F_i-4k_2z\\
G_i&\leftarrow G_i-(4k_1+k_0)z\\
F_j&\leftarrow F_j+k_2 z\\
G_j&\leftarrow G_j+k_1 z
\end{align}
where $j$ are the four neighbour sites to $i$, and the value of $f_i^{th}$ is renewed from its probability distribution. All successive unstable sites are 
treated in the same way until there are no more unstable sites. This defines the primary avalanche.
At this point the relaxation dynamics [\ref{fdot}] begins to act,  until some site eventually becomes unstable. Note that due to the discrete pinning potential, in this stage $h$ remains constant, namely the relaxation dynamics is simply:
\begin{equation}
\eta_u \partial_t F_i =-k_2 F_i,
\label{fdot2}  
\end{equation}
This means that a given site $i$ will trigger an avalanche due to relaxation if for some increase in time $\delta t$ the total force from the springs on this site, here $F_i+G_i$, reaches $f_i^{th}$, i.e., if
\begin{equation}
F_ie^{\frac{-k_2 \delta t}{\eta_u}} +G_i=f_i^{th},
\label{dt}  
\end{equation}
(note that in order to have a solution, $F_i$ must be negative, as the l.h.s. is lower that the r.h.s. at the starting time).
This leads to the determination of $\delta t$ as
\begin{equation}
\delta t =- \frac{\eta_u}{k_2}\min_i \left [\ln \left(\frac{f_i^{th}-G_i}{F_i}   \right )  \right ]
\label{dtint}
\end{equation}
Once all the secondary avalanches generated by relaxation have been produced and  $F_i$ has relaxed to zero, the external driving is increased again, according to [\ref{dtext}].

This is the main scheme of the simulation. We should mention however, that its efficient implementation relies on a classification scheme of all sites, in such a way that the determination of the next unstable site in [\ref{dtext}] and [\ref{dtint}] does not require a time consuming sweep over the whole lattice. In fact, following Grassberger \cite{Grassberger1994a} we classify the sites according to their value of the r.h.s. of [\ref{dtext}] and [\ref{dtint}], and bin them, in such a way that the determination of the next unstable site can be limited to the bin corresponding to the lowest values of these quantities. When sites change their $h$ values along the simulation, they are reaccommodated in the bins using a linked list algorithm. 
From this efficient algorithm, we obtain distributions $N(S)$ of the events sizes for various values of $k_0$, as shown in Fig. \ref{ndes} of the Supplemental Material. A power law with exponent $\kappa\simeq 1.75$ is consistently obtained.

The different dynamics we observe in 2D (power law distribution of avalanches) compared with the mean field case (periodic system size avalanches) is remarkable. From a theoretical point of view an interesting question is to identify the upper critical dimension $d_{uc}$ of the problem, beyond which its behavior is well described by the mean field results. Our qualitatively different results in 2D compared to mean field indicate that $d_{uc} >2$. 
Note that Marchetti and Dahmen proposed a mapping of the hysteretic visco-elastic model onto the
non-equilibrium random Field Ising model (RFIM) \cite{Marchetti2002b}.
This mapping seems to
suggest an upper critical dimension $d_{uc} = 6$. However there is no strong evidence for it yet.

\begin{figure}
\includegraphics{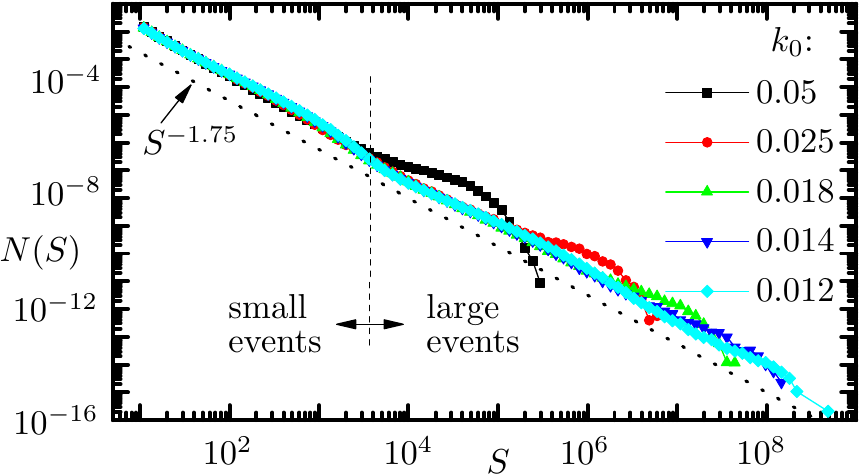}
\caption{
The size distribution of avalanches $N(S)$ is consistent with the exponent $\kappa=1.75$. 
The dotted line separating the regions of small and large events is indicated: the size distribution does not display any strong feature around this value.
The system size is in all cases much larger than the largest avalanche observed, and reaches values of $15000 \times 15000$. Other parameters are: $k_1=0$, $k_2=1$.
}
\label{ndes}
\end{figure}

One may have noticed that in the our viscoelastic model, the relaxation of the variable $u_i$
 is local and controlled by a single time constant.
This choice allows the fast computation we just described, but yields an unrealistic exponential decay of the  aftershocks over time, which does not follow the Omori law.
In this respect, it would be suitable to consider non-local relaxation mechanisms, as the Laplacian relaxation studied in  \cite{Jagla2010, Jagla2010a}, which can reproduce the Omori law.

\section{5. Parameters of the simulations}

For Figs. 1, 3 and 4, the numerical method used is the one described in the Supplemental Material (SM) Section 4. 
In Fig. 1 the number of blocks is $512^2=65536$ (in 2D, we consider a square-shaped interface with $L=512$). 
We use $k_0=0.02, k_1=0.5, g(z)=e^{-z}$ and the thresholds $f_i^{\text{th}}$ are distributed as a Gaussian of mean $3$ and unit variance.
In Fig. 1c we use $k_2=0$, i.e. we simulate the conventional depinning model.
In Fig. 1d and 1e we use $k_2=0.5$, and the grey dotted and dashed lines represent the stress averaged in small patches of size $10\times 10$ (100 blocks). The two patches are chosen to be as far as possible.

In Fig. 2 we perform the simulation described in SM Section 4 (and in particular in Sec. 4.2). 
We use $k_0 = 0.001, k_1 = 0.1, k_2 = 0.3, g(z)=e^{-z}$ and the thresholds are set constant  $f_i^{\text{th}}=1, \forall i$. The discretization of $\delta^F$ and $\delta^C$ is made using a binning $\varepsilon = 0.003$.

In Fig. 3 and 4 of the main text and Fig. 2 of the SM, we use $k_1=0, k_2=1$. $g(z)$ is the uniform distribution in the range $[0,0.2]$. The $f_i^{\text{th}}$ are distributed as in Fig. 1. The  number of blocks is $15000^2 = 225000000$ (with $L=15000$). 
In Fig. 4 we used $k_0=0.012$.
%


\end{document}